\documentclass[12pt]{iopart}
\usepackage{iopams}  
\begin{document}

\title[Deformations of WZW models]{Deformations of WZW-models}

\author{Stefan F\"orste\dag\  
\footnote[3]{email: forste@th.physik.uni-bonn.de}
}

\address{\dag\ Physikalisches Institut der Universit\"at Bonn, Nussalle 12,
  53115 Bonn, Germany }

\begin{abstract}
Current-current deformations for WZW models of semisimple compact
groups are discussed in a sigma model approach. We start with the
abelian rank one group U(1). Afterwards, we keep the rank one but
allow for non abelian structures by considering SU(2). Finally, we
present the general case of rank larger than one.

\end{abstract}



\maketitle

\section{Introduction}

Often, a consistent string background belongs to a ``family'' of
consistent backgrounds which can be deformed into each other by
continously changing parameters. A simple example is the bosonic
string compactified on an $n$ dimensional torus $T^n = U(1)^n$. In
this case the parameters which can be continuously changed are the
constant background metric and $B$ field, which can be put into an
$n\times n$ matrix $G+B$. Out of these, backgrounds are equivalent if
they are related by a T-duality transformation. The dualities are given
by the automorphisms $O(n,n,{\mathbb Z})$ of the even selfdual charge
lattice.

The subject of the present talk is to consider $U(1)^n \subset G$ where
$G$ is a semisimple compact group. The deformation of the Cartan torus
is a non-trivial modification of the discussion above since the
geometry is not a product geometry. As a warm up we will consider the
trivial case of a circle compactified string, i.e.\ $G=U(1)$. Next, we
will move to the non-abelian rank one group $SU(2)$ and finally the
general case of a semisimple compact otherwise arbitrary group will be
presented. This talk is based on the paper\cite{Forste:2003km}.

\section{Rank one, dimension one: $U(1)_R$}

The worldsheet action for a circle compactified direction $X$ is given
by
\begin{equation}
S= \frac{1}{2\pi\alpha^\prime} \int d^2z \, R^2 \partial_z X
\partial_{\bar{z}} X ,
\end{equation}
where $X$ is compactified on a unit circle and the radius of the
actual circle is encoded in the target space metric $G_{XX}= R^2$. The
corresponding equation of motion can be given the interpretation of a
chiral or an anti-chiral conversation law
\begin{equation}
\partial_{\bar{z}} J = \partial_z \bar{J}=0, \mbox{\ with\ } J =
\partial_z X, \mbox{\ and\ } \bar{J} = \partial_{\bar{z}}X .
\end{equation}
These currents are conformal primaries of dimension $(1,0)$ and
$(0,1)$, respectively. Hence the product $J\bar{J}$ is a marginal
  operator. It is easy to see that an infinitesimal change in the
  ``deformation parameter'' $R$ corresponds to a marginal perturbation,
\begin{equation}
\delta S = \frac{R\delta R}{\pi \alpha^\prime} \int d^2z J\bar{J},
\end{equation}
and hence all circle compactifications can be obtained from the string
compactified on a unit circle and exact marginal deformations. As we
will see, the situation in less trivial cases is slightly more
complicated as the form of the chiral and anti-chiral currents can
depend on the value of the deformation parameter. 

In the reminder of this section we will discuss some constructions
which are trivial in this simple case but turn out to be useful in
more general cases to be discussed later. First, we formulate the
circle compactified string as a coset theory $\left( U(1) \times
U(1)_{\tilde{R}}\right)/ U(1)$. (In later applications the first
$U(1)$ factor will be a subgroup of $G$ whose size can not be chosen
arbitrarily in a straightforward way.) The sigma model action for the
product group is
\begin{equation}
S=\frac{1}{2\pi\alpha^\prime} \int d^2 z \left( \partial_z X
\partial_{\bar{z}} X + \tilde{R}^2 \partial_z Y \partial_{\bar{z}}
Y\right) .
\end{equation}
The coset action is obtained by first gauging the isometry $X \to X +
c$ and $Y \to Y+c$. That means, that we promote the global symmetry to
a local one by replacing partial derivatives with covariant ones,
\begin{equation}
\partial_i X \to \partial_i X + A_i\,\,\, ,\,\,\, \partial_i Y \to
\partial_i Y + A_i ,
\end{equation}
where the gauge field $A_i$ transforms as $A_i \to A_i - \partial_i
c$. Next, we gauge fix e.g.\ $Y=0$ and eliminate the gauge field by
solving its algebraic equation of motion. This results in
\begin{equation}
S = \frac{1}{2\pi\alpha^\prime} \int d^2z\, R^2 \partial_z X
\partial_{\bar{z}} X ,
\end{equation}
with $R^2 = \tilde{R}^2/\left( 1 + \tilde{R}^2\right)$. Hence, we
obtain all circle compactified strings with $R\in (0,1)$. Radii larger
than one can be generated by T-duality. The unit circle, however, can
be reached only as a limit in this construction. Therefore, we will
focus 
on a different construction in the rest of the talk, namely the
orbifold construction. Let $k$ be a positive integer. Than the circle
compactified models can be written as
\begin{equation}
\left( \frac{U(1)_k}{U(1)} \times U(1)_{kR} \right)/{\mathbb Z}_k =
U(1)_{kR} /{\mathbb Z}_k = U(1)_R ,
\end{equation}
where the ${\mathbb Z}_k$ action is chosen such that it reduces the
size of the circle by  a factor $1/k$.

\section{Rank one, dimension three: SU(2)}

Next, we keep the rank of the group to be one but increase its
dimensionality, i.e.\ we consider strings on an SU(2) group
manifold. The worldsheet action for strings on group manifolds is the
WZW model action
\begin{equation}
S^{\mbox{\tiny WZW}} = S^{\mbox{\tiny kin}} +
  S^{\mbox{\tiny WZ}} = 
  \frac{k}{4\pi} 
  \left[ \int_{\Sigma} 
  L^{\mbox{\tiny kin}} +
  \int_B \omega^{\mbox{\tiny WZ}}\right], \label{WZW-model}
\end{equation}
where the level $k$ is a positive integer, $B$ is an auxiliary three
dimensional manifold whose boundary is the worldsheet $\Sigma$ and
\begin{equation}
\omega^{\mbox{\tiny WZ}} = \frac{1}{3} \mbox{Tr} \left(
g^{-1}dg\right)^3\,\,\, ,\,\,\, L^{\mbox{\tiny kin}} = \mbox{Tr}\left(
\partial_z g \partial_{\bar{z}} 
  g^{-1}\right)  \equiv -\left< g^{-1} \partial_z g ,g^{-1} \partial_{\bar{z}}
  g\right> .
\end{equation}
In order to be specific, we parameterize the SU(2) group element by
Euler angles,
\begin{equation}
g  =  \cos \! x \cos\! \tilde{\theta} -i \sin\!
    x \sin\! \theta\,\sigma^1 +  
i\sin\! x \cos\! \theta\,\sigma^2 + i\cos\! x
\sin\!\tilde{\theta}\, \sigma^3 ,
\end{equation}
such that the action (\ref{WZW-model}) now reads
\begin{eqnarray}
S^{\mbox{\tiny WZW}} & = & \frac{k}{2\pi} \int d^2
z\Big\{ \partial_z 
x\partial_{\bar{z}} x + \sin^2 x\, \partial_z \theta
\partial_{\bar{z}} \theta + \cos^2 x\, \partial_z
\tilde{\theta} 
\partial_{\bar{z}}\tilde{\theta} +\nonumber\\
& & \,\,\,\,\,\, + \cos^2x \left( \partial_z \theta \partial_{\bar{z}}  
\tilde{\theta} - \partial_z \tilde{\theta}\partial_{\bar{z}}\theta\right)
\Big\} . \label{su2}
\end{eqnarray}
Since, the metric and B-field do not depend on two of the coordinates
there is an $O(2,2,{\mathbb R})$ group generating new conformal
backgrounds. An $O(1,1,{\mathbb R})$ subgroup corresponds to exact
marginal deformations\cite{Hassan:1992gi,Giveon:1993ph}. (For general
backgrounds not depending on a certain number of coordinates the
discussion can be found in\cite{Henningson:1992rn}, for an algebraic
discussion see\cite{Freericks:1988zg,Halpern:1995js}.) Exact marginal
deformations of the SU(2) model (or its non-compact version) have been
also constructed using the coset method in\cite{Sfetsos:1993ka}. In
either of these papers one can find the action for the deformed model
to be,
\begin{eqnarray}
\hspace{-30pt}S^R & = & \frac{k}{2\pi} \int d^2z \left\{ \partial_+
  x\partial_- x + 
  \frac{\sin^2 x}{\cos^2 x + R^2 \sin^2 x} \partial_+ \theta
  \partial_- \theta   \right. \nonumber \\
& & 
\left. \hspace*{-0.5in}+\frac{ R^2 \cos^2 x}{\cos^2 x + R^2 \sin^2 
  x} 
  \partial_+  
  \tilde{\theta} \partial_- \tilde{\theta}+\frac{\cos^2 x}{\cos^2 x
  + R^2 \sin^2 x} \left( 
  \partial_+ 
  \theta \partial_- \tilde{\theta} - \partial_+
  \tilde{\theta}\partial_- \theta \right)\right\} . \label{su2defo}
\end{eqnarray}
The deformation parameter $R$ runs from zero to infinity and the
undeformed model (\ref{su2}) is obtained for $R=1$. In addition, there
is a non-trivial dilaton which will be discussed later. To confirm
that 
changing $R$ corresponds to an exact marginal deformation we first
notice that due to the equations of motion the following chiral and
anti-chiral currents are conserved:
\begin{equation}
J= k\frac{\sin^2\! x\, \partial_z \theta - \cos^2\! x\, \partial_z
  \tilde{\theta}}{\cos^2\!x + R^2 \sin^2\! x} \,\,\, ,\,\,\, 
\bar{J} = k\frac{\sin^2\! x\, \partial_{\bar{z}} \theta + \cos^2\! x\,
  \partial_{\bar{z}} 
  \tilde{\theta}}{\cos^2\!x + R^2 \sin^2\! x} .
\end{equation}
Now, it is easy to see that an infinitesimal change in the deformation
parameter corresponds to a marginal current-current perturbation,
\begin{equation}
S^{R+\delta R} = S^R - \frac{k}{2\pi} \delta R ^2 \int d^2z\, J\bar{J}
.
\end{equation}
From an algebraic perspective it has been argued that this class of
deformed SU(2) models can be described as an
orbifold\cite{Gepner:1986hr,Yang:1988bi,Giveon:1993ph},
\begin{equation} \label{su2orbi}
\mbox{Deformed Model} = \left( \frac{SU(2)_k}{U(1)}\times
U(1)_{\sqrt{k}R}\right) /{\mathbb Z}_k .
\end{equation}
However, this is also easy to see by T-dualizing the sigma model for
the orbifold\cite{Forste:2003km}. The model within the bracket of
(\ref{su2orbi}) has the action
\begin{equation}
S = \frac{k}{2\pi} \int d^2 z\left( \partial_z x\partial_{\bar{z}} x +
\tan^2\!x \, \partial_z\theta\partial_{\bar{z}}\theta + \frac{1}{R^2}
\partial_z y \partial_{\bar{z}} y\right) ,
\end{equation}
where $x$ and $\theta$ are coordinates on the coset SU(2)/U(1) whereas
$y$ is compactified on a unit circle. (Further, we have employed that
$R$ and $1/R$ are related by duality.) Now, we T-dualize the $\theta +
y$ direction, and in doing so we will incorporate the ${\mathbb Z}_k$
orbifold below. The first step in performing the T-duality is to gauge
constant shifts in $\theta + y$, i.e.\ to introduce a gauge field $A$
and replace partial derivatives with covariant ones,
\begin{equation}
\partial_i \theta \to \partial_i\theta +A_i/2 \,\,\, ,\,\,\,
\partial_i y \to \partial_i y + A_i/2 .
\end{equation}
Next, we constrain the gauge field to be locally pure gauge $A_i
=\partial_i \phi$ by adding a Lagrange multiplier term
\begin{equation} \label{constr}
\frac{1}{2\pi} \int d^2z \lambda F_{z\bar{z}} 
\end{equation}
to the action, where $F_{z\bar{z}}$ is the field strength of the gauge
field. In order to be able to absorb this pure gauge into a field
redefinition of $\theta$ and $y$ we have to worry about the global
properties of $\theta +y$ and $\phi$. Following\cite{Rocek:1991ps}, we
add a topological term
\begin{equation} \label{top}
S^{\mbox{\tiny top}} = -\frac{1}{2\pi} \int d\left( \lambda A\right) 
\end{equation}
and specify that $\lambda$ is compactified on a circle with radius
$k$, $\lambda \equiv \lambda + 2\pi k$. For a torus worldsheet the
topological term takes the form
%
$S^{\mbox{\tiny top}} = k n_a \oint_b A + k n_b \oint_a A$,
%
where $n_{a,b}$ are the winding numbers of $\lambda$ around the two
cycles of the torus labeled by $a$ and $b$. Summing over these winding
modes results in the constraint that $\phi$ is compactified on a
circle of radius $1/k$. In order, to be able to absorb $\phi$ in a
field redefinition we have to shrink the size of the $\theta + y$
circle in the product model. This is exactly what the orbifoldgroup
${\mathbb Z}_k$ in (\ref{su2orbi}) does. Finally, integrating out the
gauge field instead of $\lambda$ provides the T-dual model which is
found to be (\ref{su2defo}) after replacing $\lambda =
k\tilde{\theta}$. 

\section{Semisimple compact groups}

In this section we sketch the orbifold construction for the general
case of a semisimple compact group $G$. More details and explicit
formul\ae\ can be found in\cite{Forste:2003km}. Again, from algebraic
considerations it can be argued that the deformed models are given by
an orbifold\cite{Gepner:1987sm},
\begin{equation} \label{genorb}
\mbox{Deformed Model} = \left( G/H \times U(1)^r _E\right) /\Gamma_k ,
\end{equation}
where $H$ denotes the Cartan subgroup, $r$ the rank of the group $G$
and $E=G+B$ is the constant background on the $r$-dimensional torus
$U(1)^r$. The orbifold group is
\begin{equation}
\Gamma_k = \left( \mbox{Weight Lattice}\right)/k\left(\mbox{Lattice of
  Long Roots}\right) . \label{orbigroup}
\end{equation}
The action on a representative $g$ of the coset $G/H$ for example is
\begin{equation} \label{orbiac}
\Gamma_k: g \to \exp\left( i\vec{\varphi} \vec{H}/k\right) g,
\end{equation}
where $\vec{\varphi}$ takes values in the weight lattice. The set of
Cartan generators can be diagonalized with the eigenvalues being the
weights $\vec{\mu}_i$ ($i=1,\ldots, d-r$) ($d$ is the dimension of
$G$). If $\vec{\varphi}$ is $k$ times a long root the action is
trivial since the long root lattice is dual to the weight lattice.
The notation (\ref{orbigroup}) can be thought of as the analog of
defining ${\mathbb Z}_k$ as ${\mathbb Z}$ mod $k$, whereas
(\ref{orbiac}) is the analog of generating ${\mathbb Z}_k$ by the
$k^{\mbox{\tiny th}}$ root of unity. If we choose the generators of
$U(1)^r$ to be the same as the Cartan generators of $G$ the orbifold
acts on $U(1)^r$ as in (\ref{orbiac}) with $g$ replaced by a $U(1)^r$
element. 

In order to confirm these statements on a sigma model level, we first
take for $G/H$ the vectorially gauged WZW model. That is, for $h$
being an element of the Cartan subgroup, we promote
the global symmetry $g \to hgh^{-1}$ of (\ref{WZW-model}) to a local
one by
introducing gauge fields $A$ transforming as $A_i \to A_i +
h\partial_i h^{-1}$. Next, we add the WZW model action for $U(1)^r$
where, as described above, it is convenient to generate a $U(1)^r$
element $y$ by the Cartan generators of $G$. For abelian groups the
WZW model does not contain the WZ term. On the $U(1)^r$ model we
allow, however, for a general background such that its action is
\begin{equation}
S^{U(1)^r} = -\frac{k}{4\pi} \int d^2 z \, \langle y^{-1}\partial_z y,
E y^{-1}\partial_{\bar{z}} y\rangle ,
\end{equation}
with $E$ being a non-degenerate $r\times r$ constant matrix. The next
step is to perform a T-duality in this product and implement during
this process also the orbifold. The global symmetry with respect to
which we T-dualize is
\begin{equation} \label{tsym}
g \to fgf \,\,\, ,\,\,\, y \to f^2 y,
\end{equation}
where $f$ is an element of the Cartan subgroup of $G$. This global
symmetry is gauged in terms of a gauge field $B$ transforming as $B_i
\to B_i + f^{-1}\partial_i f$. It turns out that gauging the symmetry
(\ref{tsym}) destroys the local gauge invariance under $g \to
  hgh^{-1}$. (In asymmetrically gauged WZW models, usually a
  constraint 
  relating the two gauge fields is imposed\cite{Bars:1991pt}.) Here,
  however, we want to perform a T-duality and have to add a Lagrange
  multiplier term and a topological term (cf.\ (\ref{constr}) and
  (\ref{top})). Assigning suitable transformation properties to the
  Lagrange multipliers repairs the vector gauge invariance. The global
  properties of the Lagrange multipliers are chosen such that they are
  compactified with respect to the long root lattice. Summing over the
  corresponding winding modes specifies the global properties of the
  gauge group with elements $f$ such that a pure gauge can be absorbed
  in a field redefinition if the starting model is the orbifold
  (\ref{genorb}). Finally, integrating out the gauge fields $B$ (gauge
  fix e.g.\ $y=1$) and the gauge field $A$ (gauge fix e.g.\ $\lambda
  =0$) yields the T-dual sigma model
\begin{equation}
S = S^{\mbox{\tiny WZW}} + \frac{k}{2\pi} \int d^2 z\, \langle \left(
\mbox{PAd}_g - R^{-1}\right)^{-1} \mbox{P} \partial_z g g^{-1} ,
\mbox{P} g^{-1}\partial_{\bar{z}} g\rangle , \label{gendefo}
\end{equation}
where P denotes the projector on the Cartan subalgebra, Ad$_g$ the
adjoint action with $g$, and $R = \left(E^T - \mbox{P}\right)/\left(
E^T + \mbox{P}\right)$. We observe that the bi-invariant metric and
B-field are deformed. Hence, the original $G\times G$
chiral/anti-chiral symmetry of the WZW model is broken to
$U(1)^r\times U(1)^r$. Indeed, from (\ref{gendefo}) one finds the
following conserved currents:
\begin{eqnarray}
J& = & kR^{-1}\left( 1- RR^T\right) \left( \mbox{PAd}_g -
R^{-1}\right)^{-1} \mbox{P} \partial_z g g^{-1} ,\\
\bar{J}& = & -kR^{-T}\left( 1- R^TR\right) \left( \mbox{PAd}_{g^{-1}} -
R^{-T}\right)^{-1} \mbox{P}  g^{-1} \partial_{\bar{z}} g
\end{eqnarray}
An infinitesimal change in the deformation parameter changes the
action by 
\begin{equation}
\delta S = \frac{1}{2\pi k} \int d^2z \, \langle R\left( R^TR
-1\right)^{-1} \left( \delta R^{-1}\right) \left( 1 - RR^T\right)^{-1}
R J, \bar{J}\rangle ,
\end{equation}
which corresponds to a marginal current-current operator.

We should also remark that these models have been obtained in a coset
description for symmetric $E$ in\cite{Tseytlin:1993hm} and for general
$E$ in\cite{Forste:2003km}, again with the restriction that the
undeformed model is contained only as a limiting case. 
To conclude this section we discuss the non-trivial dilaton present in
all deformed models. There are several arguments for the source of a
non-trivial dilaton. Integrating out gauge fields means solving
Gaussian integrals which provide a determinant. On the other hand,
when 
integrating up a marginal perturbation to an exact marginal
deformation one should change path integral measures such that they
are covariant with respect to the deformed background. In any case,
the beta function equations are solved only with a non-trivial dilaton
and, indeed, this is one way to compute its form. A more elegant
prescription is given in\cite{Tseytlin:1993my}. The idea is to compare
the physical state condition that the Hamiltonian minus some normal
ordering constant should annihilate ground states with the wave
equation of the corresponding effective target space field. The wave
operator is a second order differential operator on G 
depending on the target space metric and the dilaton. The worldsheet
Hamiltonian on the other hand can be expressed in terms of affine
currents\cite{Bardakci:1970nb}, which act on ground state wave
functions as left- and 
right-invariant vector fields\cite{Bowcock:xr}. Comparing the two
operators acting 
on the ground state wave functions allows to determine the dilaton.
The result is (for details see\cite{Forste:2003km}) that
$e^{-2\Phi}\sqrt{G}$ does not change under the deformation. (Here,
$\Phi$ denotes the dilaton and $G$ the determinant of the target space
metric.)

\section{Conclusions}

We have derived the metric, B-field and dilaton of current-current
deformed WZW models for a general semisimple compact group. As in the
torus case the deformation can be viewed as deforming an even selfdual
charge lattice\cite{Forste:2003km}. Hence, the moduli space is: 
$ \mbox{Duality}\backslash O\left( r,r\right)/ O(r) \times O(r)$.
Since there are additional structures in the model the duality group
is the intersection of automorphisms for even selfdual lattices
$O\left(r,r, {\mathbb Z}\right)$ with the selfduality group of the
undeformed WZW model which is $\hat{W}\times \hat{W} \ltimes
\mbox{outer automorphisms}$, where $\hat{W}$ denotes the affine Weyl
group\cite{Kiritsis:1993ju}. 

For the future, it is planed to add D-branes which has been studied
for the SU(2) case in\cite{Forste:2001gn}. Another interesting issue
might be to add orientifolds. The action is invariant under worldsheet
parity reversal if this is combined with $ g \to c g^{-1}$, where $c$
is in the center of $G$, and $E\to E^T$. Therefore, in orientifolds $E$
has to be symmetric modulo a duality transformation, i.e.\ the B field
is quantized. As in the flat case it should be interesting to study
this phenomenon on group manifolds.
Further possible extensions are to non-semisimple and non-compact
groups.

\section*{Acknowledgments}
I would like to thank the organizers for the kind invitation to
the conference. Further, I thank Daniel Roggenkamp for collaborating
with me on the issues discussed in this talk.
My work is supported in part by the European
Community's Human Potential 
Programme under contracts HPRN--CT--2000--00131 Quantum Spacetime,
HPRN--CT--2000--00148 Physics Across the Present Energy Frontier
and HPRN--CT--2000--00152 Supersymmetry and the Early Universe, and
INTAS 00-561.


\begin{thebibliography}{99}
\bibitem{Forste:2003km}
S.~F\"orste and D.~Roggenkamp,
JHEP {\bf 0305} (2003) 071
[arXiv:hep-th/0304234].
%
\bibitem{Hassan:1992gi}
S.~F.~Hassan and A.~Sen,
Nucl.\ Phys.\ B {\bf 405} (1993) 143
[arXiv:hep-th/9210121].
%
\bibitem{Giveon:1993ph}
A.~Giveon and E.~Kiritsis,
Nucl.\ Phys.\ B {\bf 411} (1994) 487
[arXiv:hep-th/9303016].
%
\bibitem{Henningson:1992rn}
M.~Henningson and C.~R.~Nappi,
Phys.\ Rev.\ D {\bf 48} (1993) 861
[arXiv:hep-th/9301005].
%
\bibitem{Freericks:1988zg}
J.~K.~Freericks and M.~B.~Halpern,
Annals Phys.\  {\bf 188} (1988) 258
[Erratum-ibid.\  {\bf 190} (1989) 212].
%
\bibitem{Halpern:1995js}
M.~B.~Halpern, E.~Kiritsis, N.~A.~Obers and K.~Clubok,
Phys.\ Rept.\  {\bf 265} (1996) 1
[arXiv:hep-th/9501144].
%
\bibitem{Sfetsos:1993ka}
K.~Sfetsos and A.~A.~Tseytlin,
Phys.\ Rev.\ D {\bf 49} (1994) 2933
[arXiv:hep-th/9310159].
%
\bibitem{Gepner:1986hr}
D.~Gepner and Z.~a.~Qiu,
Nucl.\ Phys.\ B {\bf 285} (1987) 423.
%
\bibitem{Yang:1988bi}
S.~K.~Yang,
Phys.\ Lett.\ B {\bf 209} (1988) 242.
%
\bibitem{Rocek:1991ps}
M.~Ro\v{c}ek and E.~Verlinde,
Nucl.\ Phys.\ B {\bf 373} (1992) 630
[arXiv:hep-th/9110053].
%
\bibitem{Gepner:1987sm}
D.~Gepner,
Nucl.\ Phys.\ B {\bf 290} (1987) 10.
%
\bibitem{Bars:1991pt}
I.~Bars and K.~Sfetsos,
Mod.\ Phys.\ Lett.\ A {\bf 7} (1992) 1091
[arXiv:hep-th/9110054].
%
\bibitem{Tseytlin:1993hm}
A.~A.~Tseytlin,
Nucl.\ Phys.\ B {\bf 418} (1994) 173
[arXiv:hep-th/9311062].
%
\bibitem{Tseytlin:1993my}
A.~A.~Tseytlin,
Nucl.\ Phys.\ B {\bf 411} (1994) 509
[arXiv:hep-th/9302083].
%
\bibitem{Bardakci:1970nb}
K.~Bardakci and M.~B.~Halpern,
Phys.\ Rev.\ D {\bf 3} (1971) 2493.
%
\bibitem{Bowcock:xr}
P.~Bowcock,
Nucl.\ Phys.\ B {\bf 316} (1989) 80.
%
\bibitem{Kiritsis:1993ju}
E.~Kiritsis,
Nucl.\ Phys.\ B {\bf 405} (1993) 109
[arXiv:hep-th/9302033].
%
\bibitem{Forste:2001gn}
S.~F\"orste,
JHEP {\bf 0202} (2002) 022
[arXiv:hep-th/0112193],
Fortsch.\ Phys.\  {\bf 51} (2003) 708,
[arXiv:hep-th/0212199].
\end{thebibliography}
\end{document}